\documentclass[aps,prl,reprint, superscriptaddress]{revtex4-2}
\usepackage{bbold}
\usepackage{comment}
\usepackage[colorlinks, citecolor=red]{hyperref}
\usepackage{hyperref}
\usepackage[utf8]{inputenc}
\usepackage[T1]{fontenc}    %
\usepackage[english]{babel}
\usepackage[pdftex]{graphicx}
\usepackage{amsmath,amssymb,amsthm,bbm, mathtools} %
\usepackage{mathrsfs}
\usepackage{xcolor}
\usepackage[normalem]{ulem}
\usepackage{multirow}

\begin{document}

\title{Exchange-Only Spin-Orbit Qubits in Silicon and Germanium}

\author{Stefano Bosco}
\email{s.bosco@tudelft.nl}
\author{Maximilian Rimbach-Russ}

\affiliation{QuTech, Delft University of Technology, Delft, The Netherlands}

\begin{abstract}
The strong spin-orbit interaction in silicon and germanium hole quantum dots enables all-electric microwave control of single spins but is unsuited for multi-spin exchange-only qubits that rely on scalable discrete signals to suppress cross-talk and heating effects in large quantum processors. Here, we propose an  exchange-only spin-orbit qubit that utilizes spin-orbit interactions to implement qubit gates and keeps the beneficial properties of the original encoding. Our encoding is robust to significant local variability in hole spin properties and, because it operates with two degenerate states, it eliminates the need for the rotating frame, avoiding the technologically demanding constraints of fast clocks and precise signal calibration. Unlike current exchange-only qubits, which require complex multi-step sequences prone to leakage, our qubit design enables low-leakage two-qubit gates in a single step, addressing critical challenges in scaling spin qubits.
\end{abstract}

\maketitle

\paragraph{Introduction.}

\begin{figure}[t]
\centering
\includegraphics[width=0.49\textwidth]{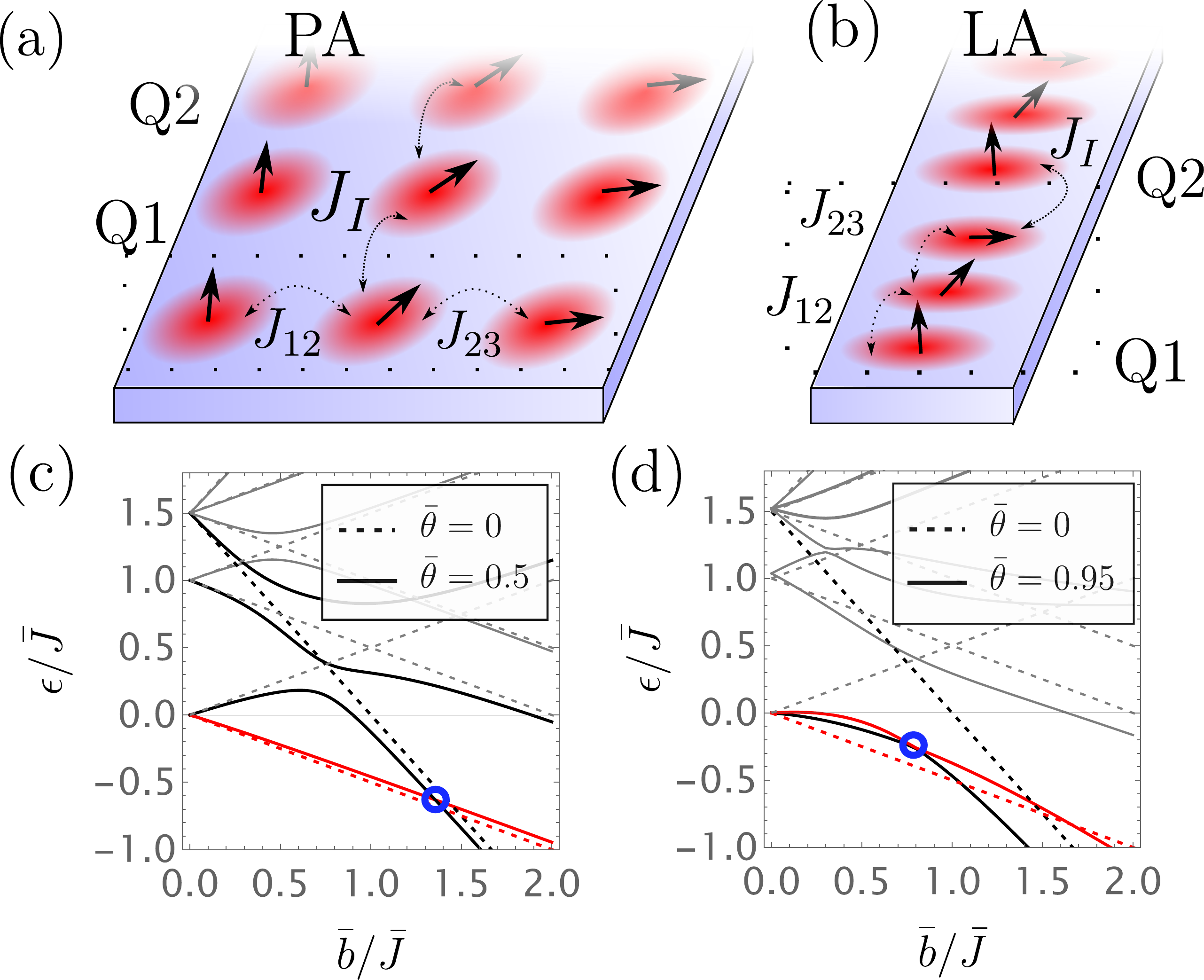}
\caption{ \textit{XOSO Qubits.} 
(a)-(b) Schematic of three quantum dots, each containing one hole spin, coupled by exchange interactions  $J_{ij}$. These interactions are anisotropic due to the strong SOI, that tilts the spin quantization axes of different dots. XOSO qubits are arranged in planar (a) and linear (b) configurations. Two-qubit gates are achieved by activating the exchange coupling $J_I$.
(c)-(d) Energy spectrum $\epsilon$ against the average Zeeman energy $\bar{b}$ shown with  (solid lines, $\bar{\theta}\neq 0$) and without SOI  (dashed lines, $\bar{\theta}=0$).  The blue circle highlights the XOSO operating point. (c) Identical quantum dots with moderate SOI $\bar{\theta}=0.5$ and no local variations of spin properties $\delta b_{ij}=\delta \theta_{ij}=\delta J=0$. (d) Locally varying quantum dots with larger SOI $\bar{\theta}=0.95$, angular variations $\delta\theta_{12}=0.3$, and varying g-factors $g_1=0.1$, $g_2=0.15$, and $g_3=0.06$. For these experimentally-relevant parameters in Si and Ge, the energy-level crossing occurs at  $\delta J=0.16\bar{J}$. 
\label{fig:1}
}
\end{figure}

Hole spin qubits in silicon (Si) and germanium (Ge) quantum dots are leading candidates for large-scale quantum computers~\cite{PhysRevA.57.120,PhysRevB.59.2070,RevModPhys.95.025003,Stano2022,Scappucci2021,annurev:/content/journals/10.1146/annurev-conmatphys-030212-184248,Fang_2023,Maurand2016, Hendrickx2021,Borsoi2024,doi:10.1126/science.ado5915,Geyer2024,zhang2023universal}. Their strong spin-orbit interaction (SOI) enables ultrafast all-electric operations~\cite{PhysRevLett.98.097202,Hendrickx2020,Froning2021,Wang2022,Camenzind2022,Jirovec2021,PhysRevLett.128.126803,liles2023singlet,Liu2023} and strong coupling to microwave photons~\cite{Yu2023,de2023strong,PhysRevLett.129.066801,PhysRevB.88.241405,PhysRevB.107.L041303,PhysRevResearch.3.013194} without requiring bulky external micromagnets. The SOI also provides a means to control qubit properties in-situ, offering sweet spots to enhance qubit performance~\cite{Piot2022,Hendrickx2024, carballido2024qubit, PRXQuantum.2.010348, Wang2021, PhysRevApplied.18.044038,wangModellingPlanarGermanium2022,PhysRevB.108.245406,PhysRevB.109.155406,PRXQuantum.5.020353,PhysRevLett.127.190501,PhysRevB.104.195421,PhysRevB.108.245406}.
To date, most spin qubits are controlled using microwave pulses~\cite{ Watzinger2018,Hendrickx20202,PhysRevLett.131.197001,PhysRevLett.132.067001,PhysRevLett.120.137702}, which introduce critical scaling challenges such as cross-talk and heating~\cite{PhysRevX.13.041015,10.1063/1.4929503}. These issues pose significant barriers to the development of large-scale devices~\cite{Vandersypen2017,PhysRevApplied.18.024053}. Some progress has been made toward discrete control of hole qubits through spin hopping in Ge quantum dots~\cite{doi:10.1126/science.ado5915,vanRiggelen-Doelman2024}, but this approach still requires the calibration of fast clocks to operate in the rotating frame, adding considerable technological constraints.

For electron-based spin qubits, the exchange-only (XO) encoding addresses these challenges by exploiting the tunability of exchange interactions between three spins confined in three quantum dots~\cite{DiVincenzo2000,PhysRevB.82.075403,gaudreauCoherentControlThreespin2012,palDrivenNonlinearDynamics2014,hungDecoherenceExchangeQubit2014,engIsotopicallyEnhancedTriplequantumdot2015,russ2017three,PhysRevLett.111.050501,PhysRevLett.111.050502,PhysRevLett.111.050503,Medford2013,shimChargenoiseinsensitiveGateOperations2016,landigCoherentSpinphotonCoupling2018,Weinstein2023,RevModPhys.95.025003,blumoffFastHighfidelityState2022,acuna2024coherent,PhysRevB.87.195309,sunFullPermutationDynamicalDecoupling2024}. This enables qubit gates through discrete pulses that switch these interactions on and off, suppressing cross-talk and and heating and eliminates the need for fast clocks. However, electron XO qubits have several drawbacks. They require more than ten pulses to implement two-qubit gates that are prone to leakage into non-computational states, causing errors that are not corrected by most quantum-error-correction algorithms~\cite{DiVincenzo2000,zeuchAnalyticPulsesequenceConstruction2014,PhysRevB.89.085314,PhysRevB.88.161303,russ2017three,salaExchangeonlySingletonlySpin2017,PhysRevLett.121.177701,jonesLogicalQubitLinear2018,andrewsQuantifyingErrorLeakage2019,panResonantExchangeOperation2020,kerckhoffMagneticGradientFluctuations2021,RevModPhys.95.025003,sunFullPermutationDynamicalDecoupling2024}. Additionally, the presence of low-lying electronic valley states disrupts the encoding~\cite{RevModPhys.85.961,PhysRevLett.88.027903,PRXQuantum.2.040358}.
In contrast, holes are not affected by small and uncontrolled valley splittings, but their large SOI hybridizes different spin sectors, leading to highly anisotropic exchange interactions~\cite{Geyer2024,zhang2023universal,saez2024microwave,PhysRevB.109.085303,PhysRevResearch.2.033036,Katsaros2020,PhysRevLett.129.116805} that introduce considerable challenges for XO qubits~\cite{gaudreauCoherentControlThreespin2012,Medford2013,engIsotopicallyEnhancedTriplequantumdot2015}.

In this work, we introduce an alternative XO encoding that leverages the strong SOI of hole spins. Unlike current XO qubits, our exchange-only spin-orbit (XOSO) qubit, depicted in Fig.~\ref{fig:1}, is encoded in the crossing of energy eigenstates with spin   $S_z=-1/2$ and  $S_z=-3/2$. This encoding is robust against significant variations in local qubit properties across the chip and removes the need for operating in the rotating frame. Remarkably, XOSO qubits not only allow full control via discrete signals due to their SOI, but they also enable two-qubit entangling gates in a single-step and are resilient to leakage during qubit operations. By mitigating key issues of hole spin  XO qubits, our XOSO encoding represents a promising path toward scalable quantum computers.

\paragraph{XOSO qubit.}

We consider a system of three hole spins arranged in a linear array of quantum dots in Si or Ge, as illustrated in Fig.~\ref{fig:1}(a)-(b). The system is modeled by the qubit-frame Hamiltonian~\cite{Geyer2024,zhang2023universal,PhysRevB.109.085303,saez2024microwave}
\begin{equation}
\label{eq:Hfull}
H= \sum_{i}\frac{b_i}{2}\sigma_z^{(i)}+ \frac{J_{12}}{4} \pmb{\sigma}^{(1)}\cdot\hat{R}_y(\theta_{12})\pmb{\sigma}^{(2)}+\frac{J_{23}}{4}\pmb{\sigma}^{(2)}\cdot\hat{R}_y(\theta_{23})\pmb{\sigma}^{(3)} \ ,
\end{equation}
which comprises Zeeman energies $b_i=g_i \mu_B B$ and anisotropic exchange interactions $J_{ij}$ between spins $i$ and $j$. The effects of site-dependent $g$-tensors and spin-flip hopping induced by large SOI are captured by the 3-dimensional rotation matrix $\hat{R}_y(\theta_{ij})$.
We introduce the average and difference of Zeeman energies $\bar{b}=\sum_i b_i/3$ and $\delta{b}_{ij}= (b_i-b_j)/2$, respectively. Similarly, for the  exchange interactions, we define $\bar{J}=(J_{12}+ J_{23})/2$, $\delta {J}=(J_{12}- J_{23})/2$, and $\bar{\theta}=(\theta_{12}+ \theta_{23})/2$, $\delta {\theta}=(\theta_{12}- \theta_{23})/2$. 

The energy spectrum for two distinct cases is shown in Fig.~\ref{fig:1}(c)-(d).
Without magnetic field ($b_i=0$), the three interacting spins form two doublets with spin $S= 1/2$ separated by $\bar{J}$ and a quartet with spin $S=3/2$ positioned at $3\bar{J}/2$ above the lowest doublet~\cite{DiVincenzo2000,PhysRevB.82.075403,russ2017three}.
As $\bar{b}$ increases, the state with spin $S_z=-3/2$ rapidly lowers its energy, eventually becoming the ground state when it crosses the lowest doublet with spin $S_z=-1/2$.
This crossing, marked by a blue circle, defines the operating point of our XOSO qubit.

For identical quantum dots and moderate SOI $\bar{\theta}$'s, the crossing occurs at $\bar{b}/\bar{J}= 3/2-23\bar{\theta}^2/40+\mathcal{O}(\bar{\theta}^4)$, see Fig.~\ref{fig:1}(c).
Although larger values of  $\bar{\theta}$ hybridize different spin sectors and significantly impacts the energy spectrum, for $\bar{\theta}\leq \pi/2$,  the crossing between the ground and first states persists albeit shifted to lower $\bar{b}/\bar{J}$ ratios. 
Remarkably, this crossing is also robust against local variations of system parameters, as shown in Fig.~\ref{fig:1}(d). Even large variations of $\delta b_{ij}$ and $\delta\theta$ are exactly compensated by  appropriately adjusting the relative exchange $\delta J$.
In experiments, $\bar{b}\gg\delta b_{ij}$ and $\bar{\theta}\gg\delta{\theta}$, allowing us to neglect the effects of local parameter variations~\cite{Geyer2024,saez2024microwave,Hendrickx2024,Jirovec2021,PhysRevLett.128.126803, liles2023singlet,zhang2023universal}. However, we emphasize that our results remain valid for more general cases with finite $\delta b_{ij}$ and $\delta\theta$, as discussed in the Supplemental Material (SM)~\cite{SM}.

\paragraph{Qubit states.}

We encode our XOSO qubit in the two-fold degenerate subspace where the ground and first excited states cross [blue circles in Fig.~\ref{fig:1}(c)-(d)]. 
Unlike current hole spin qubits and similar to XO qubits, the degeneracy of the energy states ensures that our qubit has no constant dynamics, significantly simplifying control by eliminating the need for fast clocks to synchronize qubit operations~\cite{DiVincenzo2000,shimSemiconductorinspiredDesignPrinciples2016,weissFastHighFidelityGates2022,Weinstein2023}.  
For small $\bar{\theta}$'s, our computational subspace is approximately spanned by the qubit states
\begin{subequations}
\begin{align}
|0\rangle &\approx |\downarrow\downarrow\downarrow\rangle \ , \\
|1\rangle &\approx \frac{|\uparrow\downarrow\downarrow\rangle+|\downarrow\downarrow\uparrow\rangle}{\sqrt{6}}-\sqrt{\frac{2}{3}}|\downarrow\uparrow\downarrow\rangle  \ .
\end{align}
\end{subequations}
Introducing singlet $|S_{ij}\rangle=( |\!\downarrow_{i}\uparrow_{j} \rangle - |\!\uparrow_{i}\downarrow_{j}\rangle)/\sqrt{2}$, polarized triplet $|T^\downarrow_{ij}\rangle= |\!\downarrow_{i}\downarrow_{j} \rangle $, and unpaired spin states $|\!\downarrow_{i} \rangle$ and $|\!\uparrow_{j} \rangle$  in dots  $i$ and $j$, we can rewrite the qubit states as $|0\rangle= |T_{12}^{\downarrow}\rangle|\!\downarrow_{3}\rangle = |\!\downarrow_{1}\rangle|T_{23}^{\downarrow}\rangle$ and  $|1\rangle = (|S_{12}\rangle|\!\downarrow_3~\rangle+|\!\downarrow_1\rangle|S_{23}\rangle)/\sqrt{3}-\sqrt{2/3}|T_{13}^{\downarrow}\rangle|\!\uparrow_{2}\rangle $.
This decomposition clearly shows that, in analogy to other spin qubits, our XOSO qubit can can be initialized and readout via Pauli spin blockade, which enables the distinction between singlet and triplet states in the pair of dots (1,2) or (2,3)~\cite{Takeda2024,PhysRevX.13.011023,PRXQuantum.3.040335,PhysRevB.80.041301,PhysRevB.94.041411,seedhousePauliBlockadeSilicon2021,fehseGeneralizedFastQuasiadiabatic2023,fernandez-fernandezQuantumControlHole2022a,Hendrickx2021,Lai2011,meinersenQuantumGeometricProtocols2024,doi:10.1126/science.ado5915}.

Moreover, in isotopically-purified materials, we expect that the primary source of decoherence will be  random fluctuations of the dot detuning caused by charge noise~\cite{RevModPhys.95.025003}. Although these fluctuations directly couple to the spin states, they are suppressed to first order by operating at the symmetric point with zero detuning, similar to XO qubits~\cite{russ2017three}. At this point, single- and two-qubit gates can be performed by controlling the exchange couplings by modulating the tunnel barriers between dots. We envision additional improvement by dynamical decoupling techniques tailored for XOSO qubits~\cite{PhysRevB.77.174509,rohlingOptimizingElectricallyControlled2016,West_2012,sunFullPermutationDynamicalDecoupling2024}.

\paragraph{Single-qubit gates.}

\begin{figure}[t]
\centering
\includegraphics[width=0.49\textwidth]{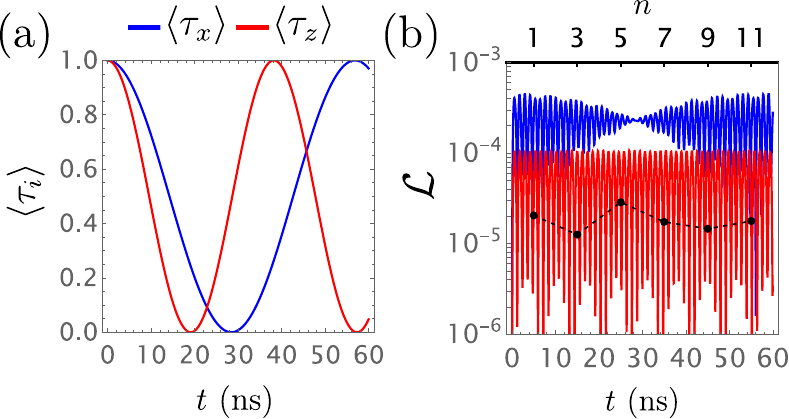}
\caption{ \textit{Single-qubit gates.}
(a) The time-dependent expectation value of $\tau_x$ ($\tau_z$) generated by applying a $\delta j/h$ ($\bar{j}/h$) pulse of $20$~MHz to the initial state $|+\rangle=(|0\rangle+|1\rangle)/\sqrt{2}$ ($|0\rangle$) is shown in blue (red) lines. We use $\bar{\theta}=0.5\approx 30^\text{o}$ and $\bar{J}/h\approx 1$~GHz. (b) Leakage $\mathcal{L}$ from the computational space. Blue and red lines correspond to the leakage over time $t$ generated by the respective single-qubit gates in~(a).  The black dots represent the leakage during the $_\varphi\text{SWAP}$ two-qubit gate, plotted against the dimensionless index $n\propto t$, corresponding to the black dots in Fig.~\ref{fig:3}(b).
\label{fig:2}
}
\end{figure}

At the degeneracy point, the SOI $\bar{\theta}$ enables the two orthogonal axes of control necessary for single-qubit operations via  global and relative exchange coupling. 
The XOSO qubit Hamiltonian, to lowest order in $\bar{\theta}$, reads
\begin{equation}
\label{eq:XOSO-SQ}
H_\text{XOSO}= -\frac{3\bar{j}(t)}{4}\tau_z-\bar{\theta}{\sqrt{\frac{3}{2}}}\frac{3  \delta j(t)}{4}\tau_x \ ,
\end{equation}
where  the Pauli matrices $\tau_i$ act on the qubit subspace, and  $\bar{ j}(t)=[J_{12}(t)+J_{23}(t)]/2-\bar{J}$ and $\delta j(t)=[J_{12}(t)-J_{23}(t)]/2 -\delta J$ represent the deviation in global and relative exchange coupling from their static values.
Higher-order corrections in $\bar{\theta}$, as well as local variations $\delta b_{ij}$ and $\delta\theta$, only quantitatively alter the driving terms, see the SM~\cite{SM}.  

We consider Si and Ge hole spin qubits, where $\bar{\theta}$ can be engineered to approach even $\pi/2$~\cite{Geyer2024}. However, because our analysis is constrained to small values of $\bar{\theta}$, we limit the discussion to moderate values $\bar{\theta}\sim 0.5\sim 30^\text{o}$. These values still enables fast electric driving, and are routinely achieved  experimentally in Ge double dots with in-plane magnetic field~\cite{saez2024microwave,Hendrickx2024}.

In Fig.~\ref{fig:2}(a), we show the time evolution of the operators $\tau_{x}$ (blue) and $\tau_z$ (red), obtained by applying a $\delta j/h$ and $\bar{j}/h$ pulse of 20~MHz. The simulation of the full $2^3$-dimensional Hamiltonian $H$ in Eq.~\eqref{eq:Hfull} aligns well with the behavior predicted by the effective XOSO Hamiltonian $H_\text{XOSO}$~\eqref{eq:XOSO-SQ}, enabling single-qubit operations in tens of nanoseconds.

\paragraph{Leakage.}

Unlike XO qubits, where leakage is a critical challenge~\cite{russ2017three,RevModPhys.95.025003}, the always-on exchange interaction $ \bar{J}$ in our XOSO qubits energetically separates the computational and non-computational subspaces, effectively suppressing leakage. 
We quantify the leakage probability during single-qubit gates by simulating the time evolution of the complete system using Eq.~\eqref{eq:Hfull} and calculating the probability of the system ending in non-computational states for initial states chosen from the six cardinal points of the Bloch sphere. Specifically, the leakage is given by $\mathcal{L}=1-\sum_{i}^N \sum_{j=0,1}|\langle j|e^{-iHt/\hbar}|i\rangle)|^2/N$, where $N=6$.  

In Fig.~\ref{fig:2}(b), we demonstrate that, under realistic parameters, leakage is substantially minimized in our system.
For an $X$ ($Z$) gate, the leakage scales as $\sim \delta j^2/\bar{J}^2$ ($\sim \bar j^2/\bar{J}^2$), meaning that the exchange pulses amplitude and the gate time are determined by $\bar{J}$.
Crucially, leakage is also suppressed during two-qubit gates. 
We straightforwardly  extended our simulation (black points) to model six interacting spins (see Fig.~\ref{fig:1}), incorporating Bell states as initial states. Because non-computational states remain gapped by $\sim \bar{J}$ even when the inter-qubit exchange interactions $J_I$ is activated, low-leakage two-qubit gates are achieved for $J_I\ll \bar{J}$.
 
Finally, we note that our simulations use simple square pulses to toggle the exchange interactions. By employing simple smoothend pulses~\cite{rimbach-russSimpleFrameworkSystematic2023}, leakage can be significantly reduced.

\paragraph{Two-qubit gates.}

\begin{figure}[t]
\centering
\includegraphics[width=0.49\textwidth]{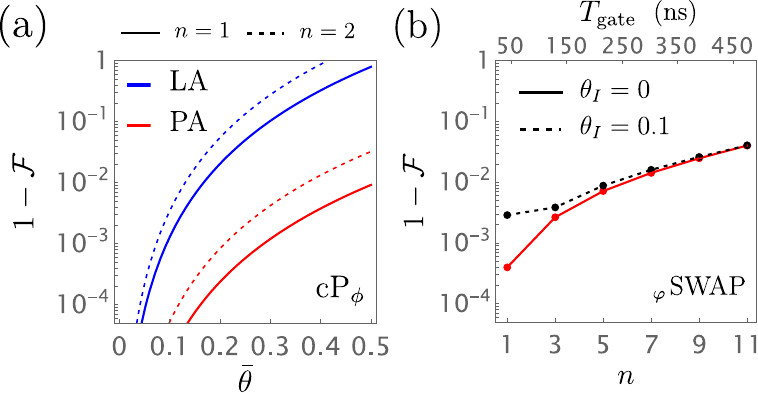}
\caption{\textit{Two-qubit gates.}
(a) Infidelity $1-\mathcal{F}$ of the $\text{cP}_\phi$ gate for the planar (PA) and linear (LA) arrangement of dots [see Fig.~\ref{fig:1}(a)-(b)] against $\bar{\theta}$. We consider a pulse duration of $t=n \tau_I/\alpha_H$ with $n=1$ (solid lines) and $n=2$ (dashed lines). Specifically, for PA (LA) with pulse amplitude $J_I/h=20$~MHz, the gate time is $T_\text{gate}= \tau_I/\alpha_H\approx 37$~ns  ($T_\text{gate}\approx 150$~ns).  (b) Infidelity of the $_\varphi\text{SWAP}$ gate  against $n$ for $\theta_I=0$ (red) and $\theta_I=0.1\approx 6^\text{o}$ (black). We simulate the full time-evolution of the six spins in PA using $\bar{J}/h=1$~GHz, $\bar{\theta}=0.5$, and $J_I/h=20$~MHz.
\label{fig:3}
}
\end{figure}

A critical advantage of our XOSO qubit is the ability to implement different high-fidelity, low-leakage two-qubit gates using a single exchange pulse, which contrasts sharply with XO qubits~\cite{DiVincenzo2000,PhysRevLett.111.050503,PhysRevB.89.085314,zeuchAnalyticPulsesequenceConstruction2014,russ2017three}.
We consider two distinct coupling schemes, the planar arrangement (PA) and linear arrangement (LA) of quantum dots, as illustrated in Fig.~\ref{fig:1}(a)-(b).
In the PA, the exchange interaction $J_I(t)$ is activated between the central dots, while in the LA, interactions are restricted to the side dots.

For both configurations,  the effective qubit-qubit Hamiltonian, to lowest order in SOI, is given by
\begin{equation}
\label{eq:H0_L}
H_{I}=\frac{J_I(t)}{8} \left[ \alpha_H
\pmb{\tau}^{(1)}\cdot \pmb{\tau}^{(2)} - \alpha_z \left({\tau}_z^{(1)}{\tau}_z^{(2)}- 
{\tau}_z^{(1)}-{\tau}_z^{(2)}\right)
\right]
 \ .
\end{equation}
This Hamiltonian contains Heisenberg-like ($\propto\pmb{\tau}^{(1)}\cdot \pmb{\tau}^{(2)}$) and Ising-like ($ \propto {\tau}_z^{(1)}{\tau}_z^{(2)}$) exchange terms, as well as single qubit terms ($\propto {\tau}_z^{(i)}$), where ${\tau}^{(i)}$ acts on qubit $i$.
The dimensionless prefactors for PA are $\alpha_H^P=4/3$, $\alpha_z^P=4/9$, while for LA  $\alpha_H^L=1/3$, $\alpha_z^L=5/18$.

Applying $J_I(t)$  for a time $t$ using a square pulse with amplitude $J_I= h/ \tau_I$  leads to the time-evolution
\begin{equation}
\label{eq:UI}
U_{I}=\left(
\begin{array}{cccc}
 1 & 0 & 0 & 0 \\
 0 & \frac{1}{2} \left(1+e^{\frac{i \pi  t \alpha _H}{\tau _I}}\right) & \frac{1}{2} \left(1-e^{\frac{i \pi  t \alpha _H}{\tau _I}}\right) & 0 \\
 0 & \frac{1}{2} \left(1-e^{\frac{i \pi  t \alpha _H}{\tau _I}}\right) & \frac{1}{2} \left(1+e^{\frac{i \pi  t \alpha _H}{\tau _I}}\right) & 0 \\
 0 & 0 & 0 & e^{\frac{i \pi  t \alpha _z}{\tau _I}} \\
\end{array}
\right) \ .
\end{equation}
When $t=T_\text{gate}= n \tau_I /\alpha_H$, with integer $n$, $U_I$ produces a family of entangling gates.
For even $n$'s, $U_I$ is diagonal and generates a controlled-phase gate $\text{cP}_\phi=\text{diag}(1,1,1, e^{i\phi})$ with $\phi=\pi n \alpha_z /\alpha_H$.
For odd $n$'s, the cP is combined with a $\text{SWAP}$ gate, i.e. $U_I=\text{SWAP} \cdot \text{cP}_{\phi}$.

The gate time $T_\text{gate}$ and the phase $\phi$ depend on the coupling scheme. For PA, $\phi^P=\pi n/3$ and for LA $\phi^L=5\pi n/6$. The gate time is given by $T_\text{gate}^P=4T_\text{gate}^L= 3n \tau_I/4$. 
Remarkably, for LA, when $n=6$, our XOSO qubits implement a controlled-Z gate with a single pulse, without requiring additional single-qubit rotations.
However, the PA enables gates that are four times faster. For example, using an exchange pulse with moderate amplitude $J_I/h=20$~MHz, which ensures low-leakage [see Fig.\ref{fig:2}(b)], fast two-qubit gates can be achieved in $T_\text{gate}^P/n\approx 37.5$~ns and  $T_\text{gate}^L/n\approx 150$~ns.
Interestingly, in addition to faster gates, the PA also enables higher-fidelity operations.

\paragraph{Gate fidelity.}

The SOI plays a crucial role in enabling single-qubit operations in our XOSO qubit system. However, it can also introduce systematic errors in the two-qubit gates by modifying the dynamics of the system. Specifically, a finite SOI angle leads to the small correction to the two-qubit interaction Hamiltonian
\begin{equation}
H^\text{1}_I=\frac{ J_I(t) {\theta}_I }{3\sqrt{6}} \left[ 
\alpha_x
\left({\tau}_x^{(1)}-{\tau}_x^{(2)}\right)-\alpha_\text{dm}\left(\pmb{\tau}^{(1)}\times \pmb{\tau}^{(2)}\right)\cdot \pmb{e}_y
\right] \ ,
\end{equation}
comprising Dzyaloshinskii-Moriya interactions ($\propto \alpha_\text{dm}$) and single-qubit terms ($\propto \alpha_x$); $\theta_I$ and $\bar{\theta}$ are  inter- and intra-qubit SOI angles, respectively.
For PA $\alpha_\text{dm}^P= 1 $, $\alpha_x^P= 1/2 $ and for LA  $\alpha_\text{dm}^L=7\bar{\theta}/20 \theta_I- 1/8 $, $\alpha_x^L=2\bar{\theta}/5\theta_I-5 /8 $.
 
The interaction term $\alpha_\text{dm}$ introduces additional entanglement between qubits that is not captured by the original gate evolution $U_I$. 
Although these interactions may be useful for engineering other types of gates, our current focus is on minimizing their effect to enhance the fidelity of the gate $U_I$.
Importantly, in both coupling schemes, we can mitigate the SOI-induced errors through careful quantum dot engineering~ \cite{PhysRevB.104.115425,PhysRevB.97.235422,PhysRevB.105.075308,PhysRevB.106.235408,PhysRevB.104.235303,PhysRevLett.131.097002,wang2022modelling,PhysRevB.103.125201,PhysRevB.108.245301}.
In PA, for isotropic inter-qubit interaction with $\theta_I^P=0$, the correction $H^\text{1}_I$ vanishes. This condition can be reached for example by engineering quantum dot shape and strain fields or by appropriately aligning the qubits with the SOI vector. Notably, these results holds even when the XOSO qubits have different values of $\bar{\theta}$.
In LA, we can eliminate $\alpha_\text{dm}^L$  by setting $\theta_I^L=14\bar{\theta}/5$, which can be done by tuning the dot positions. Any residual single-qubit dynamics can be compensated with synchronized pulses, such that $\delta j^{(1)}=-\delta j^{(2)}=J_I/5$. 

In addition to the effects linear in SOI, there are second-order SOI corrections that renormalizes the original interaction Hamiltonian in Eq.~\eqref{eq:H0_L} such that $\alpha_{H,z}\to  \alpha_{H,z}- \bar{\theta} ^2 \alpha_{H,z}^{(2)}$. These corrections also generate the terms  
\begin{equation}
\label{eq:H2SOI}
H^\text{2}_I=\frac{ J_I(t)\bar{\theta} ^2 }{4} \left[  \alpha_y{\tau}^{(1)}_y {\tau}^{(2)}_y+
\frac{\delta\alpha_z}{2}
\left({\tau}_z^{(1)}+{\tau}_z^{(2)}\right)
\right] \ .
\end{equation}
For PA  at $\theta_I^P=0$, we find $[\alpha_H^{(2)}]^P={437}/{450}$, $[\alpha_z^{(2)}]^P={227}/{270}$, $\alpha_y^P=1/3$, $\delta\alpha_z^P=22 /675$, and  at $\theta_I^L=14\bar{\theta}/5$ for LA,  $[\alpha_H^{(2)}]^L=1259/1800$, $[\alpha_z^{(2)}]^L=4837/5400$,
$\alpha_y^L=287/600$, $\delta\alpha_z^L=1789/2700$.

The full time-evolution operator $U$ including $H^2_I$ factorizes as $U=U_I U_2$, where $U_I$ is given in Eq.~\eqref{eq:UI}. The additional dynamics  $U_2$ includes SWAP transitions between $|01\rangle$ and $|10\rangle$ states  with frequency $\bar{\theta}^2\alpha_y/4\tau_I$  and double spin-flips transitions between $|00\rangle$ and $|11\rangle$ with frequency $\Omega/4\tau_I$, with $\Omega= \sqrt{\beta_z^2+\alpha_y^2\theta^4}$, and amplitude $\bar{\theta}^2\alpha_y/\Omega$. To simplify the notation, we reabsorb the corrections $\alpha_{H,z}^{(2)}$ into $\alpha_{H,z}$, and we introduce the prefactor  $\beta_z=\alpha_z+\bar{\theta}^2\delta \alpha_z$ of the single-particle term $\propto (\tau_z^{(1)}+\tau_z^{(2)})$.  The full form of $U$ is given in the SM~\cite{SM}.

The two-qubit gate fidelity is extracted from the explicit form of $U$  at $t=n\tau_I/\alpha_H $  as
\begin{equation}
\mathcal{F}=\left|\frac{1}{4}\text{Tr}(U_I^\dagger U)\right|^2\approx 1-C_n \bar{\theta}^4  \ ,
\end{equation}
where the dimensionless constant $C_n\sim n^2$ depends on the coupling scheme~\cite{SM}. Fig.~\ref{fig:2}(a) shows a comparison of $\mathcal{F}$ for LA and PA  across different $n$ values. The PA  not only provides faster gate times but also higher fidelity.

\paragraph{SOI-enabled two-qubit gates.}

Interestingly, $H^\text{2}_I$~\eqref{eq:H2SOI} can be leveraged to create a different class of two-qubit gates that are resilient to third-order SOI effects. By applying a pulse with $\bar{j}^{(1)}=\bar{j}^{(2)}=\bar{j}$, we can tune the single-qubit terms such that $\beta_z \to \beta_z^*=\beta_z+6\bar{j}/J_I$, and the double-spin-flip frequency $\Omega\to\Omega^*=\sqrt{(\beta_z^*)^2+\bar{\theta}^4\alpha_y^2} $. This modification enables us to synchronize the SWAP and double-spin-flip oscillations. Choosing $\bar{j}$ such that $\Omega^* t=2\tau_I$, at $t={T}_\text{gate}= n\tau_I/(\alpha_H+\bar{\theta}^2\alpha_y)$, with odd values of $n$,  results in the phase SWAP
\begin{equation}
\label{eq:Uiswap}
_{\varphi}\text{SWAP}=\left(
\begin{array}{cccc}
 1 & 0 & 0 & 0 \\
 0 &0& e^{i\varphi} & 0 \\
 0 &  e^{i\varphi} & 0 & 0 \\
 0 & 0 & 0 & 1 \\
\end{array}
\right) \ ,
\end{equation}
with phase $\varphi/\pi=1-n (\alpha_z+\alpha_y\bar\theta^2)/2(\alpha_H+\alpha_y\bar\theta^2)$. 
For PA and LA, we find $\varphi^P/\pi-1\approx - n(1/6- 199\bar{\theta}^2/1800)$ and $\varphi^L/\pi-1\approx - n(5/12-1259\bar{\theta}^2/3600)$, respectively.

We confirm numerically that this class of two-qubit gates achieves high-fidelity by simulating the dynamics of six interacting spins under the appropriate $\bar{j}$ pulse. Fig.~\ref{fig:3}(b) focuses on the PA with a pulse $J_I/h=20$~MHz and $\bar{\theta}=0.5$,  illustrating the infidelity arising from higher-order SOI corrections and finite  $\theta_I$. 
This entangling gate can be implemented at high speed, high fidelity, and low leakage, see Fig.\ref{fig:2}(b).  

We anticipate that gate fidelities can be further enhanced by utilizing various techniques.
SOI in hole systems is highly tunable via quantum dot and strain engineering~\cite{PRXQuantum.2.010348,PhysRevB.104.115425,PhysRevB.97.235422,PhysRevB.105.075308,PhysRevB.106.235408,PhysRevB.104.235303,PhysRevLett.131.097002,wang2022modelling,PhysRevB.103.125201,PhysRevB.108.245301}, enabling protocols that could dynamically switch off SOI during two-qubit gates, thus eliminating systematic errors. Additionally, pulse shaping and optimal control techniques, which were not explored here, are known to significantly improve qubit performance~\cite{rimbach-russSimpleFrameworkSystematic2023}.

\paragraph{Conclusion.}

We introduce the XOSO qubit, comprising three hole spins located in Si or Ge quantum dots. Our encoding leverages the strong SOI in holes enabling full qubit control using discrete signals and removes the need for a fast clock. This significantly reduces the technological challenges for scaling up quantum devices.
In contrast to XO qubits, our XOSO qubits are intrinsically protected against leakage during qubit operations, allowing for fast high-fidelity two-qubit entangling gates with just a single pulse. This  makes XOSO qubits a scalable and promising alternative for large-scale spin-based quantum processors.

\paragraph{Acknowledgement.}

This research was partly supported by the EU through the H2024 QLSI2 project and partly sponsored by the Army Research Office under Award Number: W911NF-23-1-0110. M R-R acknowledges additional support from NWO under Veni Grant (VI.Veni.212.223). The views and conclusions contained in this document are those of the authors and should not be interpreted as representing the official policies, either expressed or implied, of the Army Research Office or the U.S. Government. The U.S. Government is authorized to reproduce and distribute reprints for Government purposes notwithstanding any copyright notation herein.
 
\bibliography{literature}

%%%%%%%%%%%%%%%% SUPPLEMENTARY
\clearpage
\newpage
\mbox{~}
%\clearpage
%\newpage

\onecolumngrid

\begin{center}
  \textbf{\large {Supplemental Material: \\ Exchange-Only Spin-Orbit Qubits in Silicon and Germanium}\\[.2cm]}
  Stefano Bosco, Maximilian Rimbach-Russ\\[.1cm]
  {\itshape QuTech, Delft University of Technology, Delft, The Netherlands}\\
\end{center}

\setcounter{equation}{0}
\setcounter{figure}{0}
\setcounter{table}{0}
\setcounter{section}{0}

\renewcommand{\theequation}{S\arabic{equation}}
\renewcommand{\thefigure}{S\arabic{figure}}
\renewcommand{\thesection}{S\arabic{section}}
\renewcommand{\bibnumfmt}[1]{[S#1]}
\renewcommand{\citenumfont}[1]{S#1}

\section*{abstract}
In the Supplemental Material, we show the effective Hamiltonian of the Exchange-Only-Spin-Orbit (XOSO) qubit in presence of large variations of qubit parameters. We also report the complete expression of the time-evolution operator of two-interacting XOSO qubits.

\section{Effective theory for varying parameters}
Here, we derive the general effective Hamiltonian of XOSO qubits including arbitrary qubit g-factors and spin-orbit interaction (SOI) angles between the quantum dots.
We start from the Hamiltonian describing three interacting spins in three quantum dots
\begin{equation}
H= \sum_{i}\frac{b_i}{2}\sigma_z^{(i)}+ \frac{J_{12}}{4} \pmb{\sigma}^{(1)}\cdot\hat{R}_y(\theta_{12})\pmb{\sigma}^{(2)}+\frac{J_{23}}{4}\pmb{\sigma}^{(2)}\cdot\hat{R}_y(\theta_{23})\pmb{\sigma}^{(3)} \ .
\end{equation}
In analogy to the main text, we use an Hamiltonian in the qubit frame that models the effect of spin-orbit interactions and tilt of quantization axes of spin in different quantum dots by the anisotropic exchange interaction $J_{ij} R_y(\theta_{ij})$, where $R_y(\theta_{ij})$ is a 3-dimensional rotation matrix.

We introduce the average and difference of Zeeman energies $\bar{b}=\sum_i b_i/3$ and $\delta{b}_{ij}= (b_i-b_j)/2$, respectively, and exchange amplitudes $\bar{J}=(J_{12}+ J_{23})/2$, $\delta {J}=(J_{12}- J_{23})/2$, and SOI angles $\bar{\theta}=(\theta_{12}+ \theta_{23})/2$, $\delta {\theta}=(\theta_{12}- \theta_{23})/2$. 
We restrict ourselves to systems where the relative quantities are smaller than the average quantities.

Without SOI $\theta_{ij}=0$ and at $\delta{b}_{ij}=\delta J=0$, the eigenstates are two spin 1/2 doublets with energies $\pm \bar{b}/2$ and $\bar{J}\pm \bar{b}/2$, which are used for conventional XO qubits, and a spin 3/2 quartet with energies $3\bar{J}/2\pm \bar{b}/2$ and $3\bar{J}/2\pm 3\bar{b}/2$.
We focus on the crossing between the groundstate doublet at spin $S_z=-1/2$ and the quartet with spin $S_z=-3/2$, and account for the additional terms in the Hamiltonian perturbatively by applying a standard Schrieffer-Wolff transformation. 
With this procedure and up to second order, we find that the crossing occurs at 
\begin{align}
\frac{\bar{J}}{\bar{b}}&\approx\frac{2}{3}-\frac{\delta b_{12}-\delta b_{23}}{9 \bar{b}}-\frac{4(\delta b_{12}+\delta b_{23}) \tan\! {\delta \theta } \cot\! \bar\theta+2(\delta b_{12}^2+\delta b_{23}^2)}{27 \bar{b}}+\frac{280 \cos ^2\!\frac{\delta \theta }{2} \sin ^2\!\frac{\bar\theta }{2}-\cos ^2\!{\delta \theta } \sin ^2\!\bar\theta -40 \tan ^2\!{\delta \theta } \cot ^2\!\bar\theta}{270} \ , \\
\frac{\delta{J}}{\bar{b}}&\approx \frac{\delta b_{12}+\delta b_{23}}{9\bar{b}}-\frac{4}{9} \tan {\delta \theta } \cot  \bar\theta -2 \frac{\delta b_{12}^2-\delta b_{23}^2}{27 \bar{b}^2}\ .
\end{align}

Working at the crossing, we find the effective XOSO Hamiltonian including the first corrections
\begin{align}
H_\text{XOSO}=&-\frac{3 \bar{j}(t)}{4}\left(1-\frac{7 \sin ^2\!\frac{\bar\theta }{2} (17 \cos\! \bar\theta+217)}{900} -\frac{2 \delta \theta  (\delta b_{12}+\delta b_{23}) \cot\! \bar\theta}{9 \bar{b}}-\frac{\delta b_{12}^2+\delta b_{23}^2}{9 \bar b^2}
\right)\tau_z \\
&-\frac{3 \delta{j}(t)}{4}\sqrt{\frac{3}{2}}\left(\cos \delta \theta  \sin \bar\theta -\frac{  (\delta b_{12}-\delta b_{23}) \cos \delta \theta  \sin \bar\theta }{6 \bar{b}}\right)\tau_x \\
&+\frac{3 \delta{j}(t)}{4}\left(\frac{2}{3} \tan \delta \theta \cot \bar\theta
+\frac{  \delta b_{12}+\delta b_{23}}{3 \bar{b}} -\frac{  \delta b_{12}^2-\delta b_{23}^2}{6 \bar{b}^2}-\frac{2 (\delta b_{12}-\delta b_{23}) \tan \delta \theta \cot\bar\theta}{9 \bar b}
\right)\tau_z \\
& -\frac{3 \bar{j}(t)}{8}\sqrt{\frac{3}{2}}\left(\cos \bar\theta  \sin \delta\theta -\frac{  (\delta b_{12}+\delta b_{23}) \cos \delta \theta  \sin \bar\theta }{6 \bar{b}}\right)
\tau_x \ ,
\end{align}
where $\tau_i$ is the $i$th Pauli matrices of the effective low-energy two-level system. To simplify the notation, we also expanded the term in the first line for small values of $\delta\theta$.
By neglecting the corrections caused by $\delta \theta$ and $\delta b_{ij}$, and expanding in $\bar{\theta}$ to the lowest order, we obtain Eq.~(3) in the main text.

\section{General dynamics of two XOSO qubits}
We consider here the general two-qubit interaction Hamiltonian of the form
\begin{equation}
\label{eq:H0_L}
H=\frac{J_I}{8} \left[ \alpha_H \pmb{\tau}^{(1)}\cdot \pmb{\tau}^{(2)} - \alpha_z{\tau}_z^{(1)}{\tau}_z^{(2)}+ 2\alpha_y{\tau}_y^{(1)}{\tau}_y^{(2)}+ 
\beta_z({\tau}_z^{(1)}+{\tau}_z^{(2)})+ 
\delta\beta_z({\tau}_z^{(1)}-{\tau}_z^{(2)})
\right]
 \ .
\end{equation}
The resulting time-evolution operator is given by $U=e^{-i H t'/\hbar}$, and explicitly it reads, up to irrelevant global phases,
\begin{equation}
U=\left(
\begin{array}{cccc}
 \cos \left(\frac{\pi  t \Omega }{2}\right)-\frac{i \beta _z \sin \left(\frac{\pi  t \Omega }{2}\right)}{\Omega } & 0 & 0 & \frac{i \alpha _y \sin \left(\frac{\pi  t \Omega }{2}\right)}{\Omega } \\
 0 & \frac{e^{i \tilde{\varphi}} \left[i \omega  \cos \left(\frac{\pi  t \omega }{2}\right) +\delta \beta_z  \sin \left(\frac{\pi  t \omega }{2}\right)\right]}{\omega } & \frac{e^{i \tilde{\varphi}} \left(\alpha_H+\alpha _y\right) \sin \left(\frac{\pi  t \omega }{2}\right)}{\omega } & 0 \\
 0 & \frac{e^{i \tilde{\varphi}} \left(\alpha_H+\alpha _y\right) \sin \left(\frac{\pi  t \omega }{2}\right)}{\omega } & \frac{e^{i \tilde{\varphi}} \left[i \omega  \cos \left(\frac{\pi  t \omega }{2}\right) -\delta \beta_z  \sin \left(\frac{\pi  t \omega }{2}\right)\right]}{\omega } & 0 \\
 \frac{i \alpha _y \sin \left(\frac{\pi  t \Omega }{2}\right)}{\Omega } & 0 & 0 & \cos \left(\frac{\pi  t \Omega }{2}\right)+\frac{i \beta _z \sin \left(\frac{\pi  t \Omega }{2}\right)}{\Omega } \\
\end{array}
\right)\  .
\end{equation}
To simplify the notation, we scale the time by $\tau_I=h/J_I$ such that $t= t'/\tau_I$, and we introduce the rotation frequencies and phase
\begin{align}
\Omega &= \sqrt{\alpha _y^2+\beta _z^2} \ , \\ 
\omega &= \sqrt{\delta \beta_z ^2+\left(\alpha_H+\alpha _y\right){}^2} \ , \\ 
\varphi&=\tilde{\varphi}+\pi \beta_z t/2 = \frac{ \pi}{2}  \left[(\alpha_H-\alpha _z+\beta_z) t -1\right] \ .
\end{align}

In the cases discussed in the main text, where $\delta\beta_z=0$, we can conveniently decompose $U$ as  $U= U_I U_2$, with  
\begin{align}
U_I=\left(
\begin{array}{cccc}
 1 & 0 & 0 & 0 \\
 0 & i e^{i \varphi } \cos \left( \pi  \alpha _H t/2\right) & e^{i \varphi } \sin \left( \pi   \alpha _H t/2\right) & 0 \\
 0 & e^{i \varphi } \sin \left( \pi   \alpha _H t/2\right) & i e^{i \varphi } \cos \left( \pi   \alpha _H t/2\right) & 0 \\
 0 & 0 & 0 & e^{i \pi  t \beta _z} \\
\end{array}
\right) \ ,
\end{align}
obtained for $\alpha_y=0$ and 
\begin{equation}
U_2=\left(
\begin{array}{cccc}
 e^{\frac{1}{2} i \pi  t \beta _z} \left(\cos \left(\frac{\pi  t \Omega }{2}\right)-\frac{i \beta _z \sin \left(\frac{\pi  t \Omega }{2}\right)}{\Omega }\right) & 0 & 0 & \frac{i \alpha _y \sin \left(\frac{\pi  t \Omega }{2}\right) e^{\frac{1}{2} i \pi  t \beta _z}}{\Omega } \\
 0 & \cos \left( \pi  t \alpha _y/2\right) & -i \sin \left( \pi  t \alpha _y/2\right) & 0 \\
 0 & -i \sin \left(\pi  t \alpha _y/2\right) & \cos \left( \pi  t \alpha _y/2\right) & 0 \\
 \frac{i \alpha _y \sin \left(\frac{\pi  t \Omega }{2}\right) e^{-\frac{1}{2} i \pi  t \beta _z}}{\Omega } & 0 & 0 & e^{-\frac{1}{2} i \pi  t \beta _z} \left(\cos \left(\frac{\pi  t \Omega }{2}\right)+\frac{i \beta _z \sin \left(\frac{\pi  t \Omega }{2}\right)}{\Omega }\right) \\
\end{array}
\right) \ .
\end{equation}

At times $t=n/\alpha_H$, the term $U_I$ generates a controlled-phase two-qubit gate $\text{cP}_\phi$ and a controlled-phase gate with an additional SWAP, depending on the parity of the integer $n$.
The fidelity of the $\text{cP}_\phi$ two-qubit gate can be found from $\mathcal{F}=\left|\frac{1}{4}\text{Tr}(U_I^\dagger U)\right|^2=\left|\frac{1}{4}\text{Tr}(U_2)\right|^2$
\begin{equation}
\mathcal{F}=\frac{1}{4} \left[\cos \left(\frac{\pi  t \alpha _y}{2} \right)+\cos \left(\frac{\pi  t \Omega }{2}\right) \cos \left(\frac{ \pi  t \beta _z}{2}\right)+\frac{\beta _z \sin \left(\frac{\pi  t \Omega }{2}\right) \sin \left(\frac{\pi  t \beta _z}{2} \right)}{\Omega }\right]^2\approx 1-\frac{\alpha _y^2 }{8} \left(\pi ^2 t^2+\frac{4 \sin^2 \left(\frac{\pi  t \beta _z}{2}\right)}{\beta _z^2}\right) \ .
\end{equation}
By substituting  $t=n/\alpha_H$ and expanding $\mathcal{F}$ for small values of $\bar{\theta}$, we obtain the Eq.~(8) in the main text.

\end{document}